\documentclass[aps,prl,twocolumn,showpacs,preprintnumbers,amsmath,amssymb,a4paper]{revtex4}

\usepackage{graphicx}
\usepackage{dcolumn}
\usepackage{bm}

\begin{document}
\preprint{J. Y. Jo \textit{et al.}}
\title{Domain Switching Kinetics in Disordered Ferroelectric Thin Films}
\author {J. Y. Jo,$^{1}$ H. S. Han,$^{1}$ J.-G. Yoon,$^{2}$ T. K. Song,$^{3}$ S.-H. Kim,$^{4}$}
\author {T. W. Noh$^{1,}$}
\email{twnoh@snu.ac.kr}
\affiliation{$^{1}$ReCOE\,$\&$FPRD,\;Department\,of\,Physics\,and\,Astronomy,\,Seoul\,National\,University,\,Seoul\,151-747,\,Korea\break
$^{2}$Department  of Physics,University  of  Suwon, Suwon,
Gyeonggi-do 445-743, Korea \break $^{3}$School of Nano Advanced
Materials, Changwon  National University, Changwon, Gyeongnam
641-773, Korea\\$^{4}$R$\&$D center, Inostek Inc., Ansan,
Gyeonggi-do 426-901, Korea}
\date{\today}
\begin{abstract}
We investigated domain kinetics by measuring the polarization
switching behaviors of polycrystalline Pb(Zr,Ti)O$_{3}$ films,
which are widely used in ferroelectric memory devices. Their
switching behaviors at various electric fields and temperatures
could be explained by assuming the Lorentzian distribution of
domain switching times. We viewed the switching process under an
electric field as a motion of the ferroelectric domain  through a
random medium, and we showed that the local field variation due to
dipole defects at domain pinning sites could explain the
intriguing distribution.
\end{abstract}

\maketitle Domain switching kinetics in ferroelectrics (FEs) under
an external electric field $E_{ext}$ have been extensively
investigated for several decades
\cite{Scott1,YWSo,Lohse,Tagantsev,Stolichnov,Gruverman,Shur,Stolichnov2,BHPark}.
The traditional approach to explain the FE switching kinetics,
often called the Kolmogorov-Avrami-Ishibashi (KAI) model, is based
on the classical statistical theory of nucleation and unrestricted
domain growth \cite{Kolmogorov, Avrami}. For a uniformly polarized
FE sample under $E_{ext}$, the KAI model gives the time
($t$)-dependent change in polarization $\Delta P$($t$) as
\begin{equation} \textstyle \Delta
P(t)=2P_{s}[1-exp\{-(t/t_{0})^{n}\}],
\end{equation}
where $n$ and $t_{0}$ are the effective dimension and
characteristic switching time for the domain growth, respectively,
and $P_{s}$ is spontaneous polarization. When the nuclei are
appearing in time with the same probability, $n$ = 3 for bulk
samples and $n$ = 2 for thin films \cite{remark1}. In addition,
$t_{0}$ is proportional to the average distance between the
nuclei, divided by the domain wall speed. Several studies have
used the KAI model successfully to explain the $\Delta P$($t$)
behaviors of FE single crystals and epitaxial thin films
\cite{YWSo}.
\begin{figure}[b]
\includegraphics[width=2.5in]{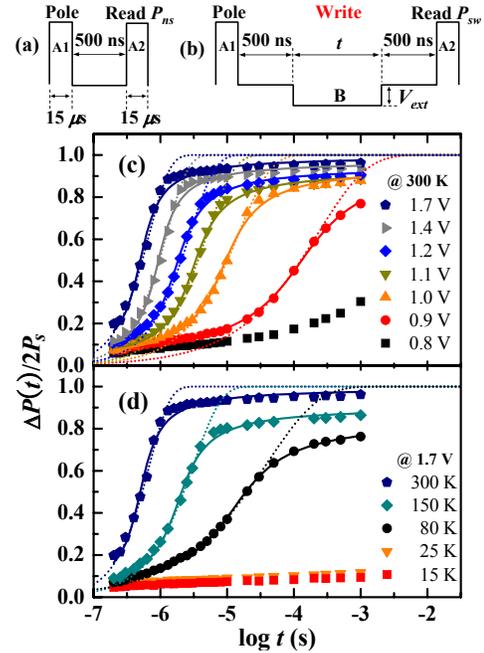}
\caption{(color online). Schematic diagrams of the pulse trains
used to measure (a) non-switching polarization ($P_{ns}$) and (b)
switching polarization ($P_{sw}$). Time ($t$)-dependent switched
polarization  $\Delta P$($t$) (c) under various external voltages
($V_{ext}$) at room temperature and (d) under 1.7 V at various
temperatures. The dotted and solid lines correspond to fitted
results using the KAI model and the Lorentzian distribution
function in log $t_{0}$, respectively.}
\end{figure}

Recently, FE thin films have been intensively investigated for FE
random access memory (FeRAM) \cite{Scott1}. Most commercial FeRAM
use polycrystalline Pb(Zr,Ti)O$_{3}$ (poly-PZT) films, and their
$\Delta P$($t$) behaviors determine the reading and writing speeds
of the FeRAM. In such non-epitaxial FE films, a domain cannot
propagate indefinitely due to pinning caused by numerous defects,
so the KAI model cannot be applied. Therefore, it is important
both scientifically and technologically to clarify the domain
switching kinetics of
polycrystalline FE films.\\
\indent Numerous studies have examined the $\Delta P(t)$ behaviors
of polycrystalline FE films, and the reported results vary
markedly \cite{Shur,Lohse,Tagantsev, Stolichnov,Gruverman}. Lohse
$et$ $al.$ measured the polarization switching currents of
poly-PZT films, and showed that $\Delta P$($t$) slowed
significantly compared to Eq. (1) \cite{Lohse}. Tagantsev $et$
$al.$ observed similar phenomena for poly-PZT films. To explain
these behaviors, they developed the nucleation-limited-switching
(NLS) model. They assumed that films consist of several areas that
have independent switching kinetics:
\begin{equation}
\textstyle \Delta
P(t)=2P_{s}\int_{-\infty}^{\infty}[1-exp\{-(t/t_{0})^{n}\}]F(log\,t_{0})d(log\,t_{0}),
\end{equation}
where $F$(log $t_{0}$) is the distribution function for log
$t_{0}$ \cite{Tagantsev}. They assumed a very broad mesa-like
function for $F$(log $t_{0}$), and could explain their $\Delta
P$($t$) data. The same authors also studied La-doped poly-PZT
films and found that $\Delta P(t)$ at room temperature is limited
mainly by nucleation, while at a low temperature ($T$), the
switching kinetics are governed by domain wall motion, implying
the validity of the KAI model \cite{Stolichnov}.

\begin{figure}[floatfix]
\includegraphics[width=2.4in]{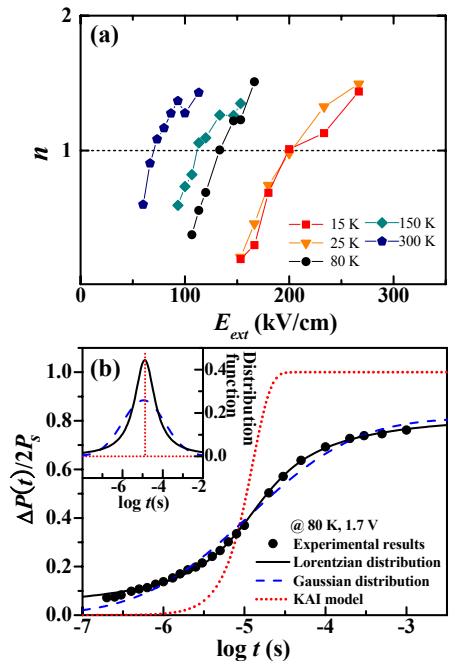}
\caption{(color online). (a) Values of $n$ for various $T$ and
$E_{ext}$. (b) $\Delta P$($t$) results for (solid symbols)
experimental data and fitted results using the Lorentzian (solid
line), Gaussian (dashed line), and delta (dotted line)
distributions for log $t_{0}$. The inset shows the distribution
functions corresponding to the fitted results.}
\end{figure}

In this Letter, we investigate the polarization switching
behaviors of poly-PZT films. We can explain the measured $\Delta
P$($t$)
 in terms of the Lorentzian distribution function for
$F$(log $t_{0}$), irrespective of $T$. We show that such
distribution arises from local field variation in a disordered
system with dipole-dipole
interactions.\\
\indent Note that (111)-oriented poly-PZT films with a Ti
concentration near 0.7 are the most widely used material in FeRAM
applications. We prepared our polycrystalline
PbZr$_{0.3}$Ti$_{0.7}$O$_{3}$ thin film on Pt/Ti/SiO$_{2}$/Si
substrates using the sol-gel method. The poly-PZT film had a
thickness of 150 nm. X-ray diffraction studies showed that it has
the (111)-preferred orientation, and scanning electron microscopy
studies indicated that our poly-PZT film consists of grains with a
size of about 200 nm. We deposited Pt top electrodes using
sputtering with a shadow mask. The areas of the top electrodes
were about 7.9$\times$10$^{-9}$ m$^{2}$.

We obtain the $\Delta P$($t$) values of our Pt/PZT/Pt capacitors
using pulse measurements \cite{Tagantsev,YWSo,YSKim,JYJo1}. Figure
1(a) shows the pulse trains used to measure the non-switching
polarization change ($P_{ns}$). Using pulse A1, we poled all the
FE domains in one direction. Then, we applied pulse A2 with the
same polarity, and measure the current passing a sensing resistor.
By integrating the current, we could obtain the $P_{ns}$ values.
Figure 1(b) shows the pulse trains used to measure the switching
polarization ($P_{sw}$). Inserting pulse B with the opposite
polarity between pulses A1 and A2, we could reverse some portion
of the FE domains, so the difference between the values of
$P_{sw}$ and $P_{ns}$ represents the polarization change due to
domain switching, namely $\Delta P$($t$). We varied $t$ from 200
ns to 1 ms, and $V_{ext}$ from 0.8 to 4 V. The value of $E_{ext}$
can be estimated easily by dividing $V_{ext}$ by the film
thickness. At $T$ of 80$\sim$300 K, we used pulses A1 and A2 with
a height of 4 V, which was larger than the coercive voltage. Below
80 K, the coercive voltage increases, so we increased
the pulse height to 6 V \cite{remark2}.\\
\indent Figure 1(c) shows the values of $\Delta P$($t$)/2$P_{s}$
at room temperature with numerous values of $V_{ext}$. Figure 1(d)
shows the values of $\Delta P$($t$)/2$P_{s}$ at various $T$ with
$V_{ext}$ = 1.7 V. The dotted lines in both figures are the curves
best fitting Eq. (1). The KAI model predictions deviated markedly
 from the experimental $\Delta P$($t$) values in the late
switching stage, in agreement with Gruverman $et$ $al$.
\cite{Gruverman}. In addition, the best fitting results with the
KAI model gave unreasonable values of $n$. As shown in Fig. 2(a),
the values of $n$ varied markedly with $T$ and $E_{ext}$. In
addition, in the low $E_{ext}$ region, we obtained $n$ values much
smaller than 1, which are not proper as an effective dimension of
domain growth. Therefore, Eq. (1) fails to describe the
polarization switching behaviors of our PZT films.
\begin{figure}[floatfix]
\includegraphics[width=2.4in]{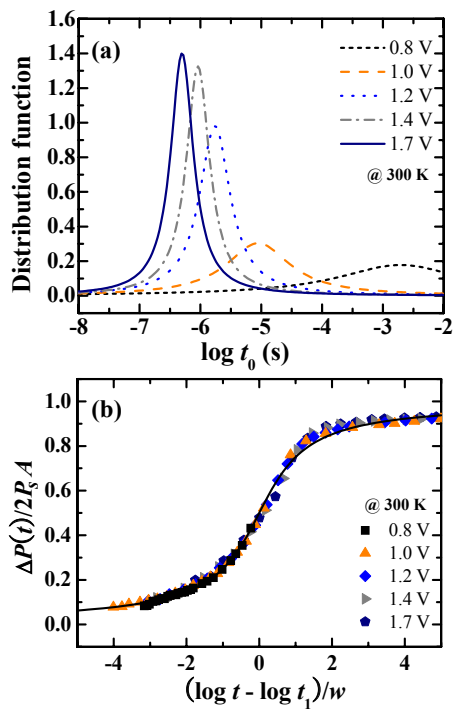}
\caption{(color online). (a) The $E_{ext}$-dependent Lorentzian
distribution functions at room temperature. (b) Rescaled $\Delta
P$($t$) using fitting parameters for the Lorentzian distribution
function.}
\end{figure}

To explain the measured $\Delta P$($t$), we tried simple functions
for $F$(log $t_{0}$) in Eq. (2). The opposite domain, once
nucleated, will propagate inside the film, so we fixed $n$=2. The
solid circles in Fig. 2(b) show the experimental $\Delta P($t$)$
at 80 K with $V_{ext}$ = 1.7 V. For $F$(log $t_{0}$), we tried the
delta, Gaussian, and Lorentzian distribution functions, as shown
in the inset. The dotted line indicates the fitting results using
Eq. (2) with a delta function. Note that this curve corresponds to
a fit with the KAI model, and thus the classical theory cannot
explain our experimental data. The dashed line shows the Gaussian
fitting results. Although this fitting seems reasonable, some
discrepancies occur. The solid black shows the fitting results
with the Lorentzian distribution:
\begin{equation}
\textstyle  F(log\,t_{0})=\frac{A}{\pi}
\left[\frac{w}{(log\,t_{0}-log\,t_{1})^{2}+w^{2}}\right],
\end{equation}
where $A$ is a normalization constant, and $w$ (log $t_{1}$) is
the half-width at half-maximum (a central value) \cite{remark3}.
The Lorentzian fit can account for our observed $\Delta P$($t$)
behaviors quite well.

We applied the Lorentzian fit to all of the other experimental
$\Delta P$($t$) data. As shown by the solid lines in Figs. 1(c)
and (d), the Lorentzian fit provides excellent explanations.
Figure 3(a) presents the Lorentzian distribution functions used
for the 300 K data. As $V_{ext}$ increases, log $t_{1}$ and $w$
decrease. We rescaled the experimental $\Delta P$($t$)/2$P_{s}$
data using (log $t$ - log $t_{1}$)/$w$. All the data merge into a
single line, an arctangent function \cite{remark3}, as shown in
Fig. 3(b). Although not indicated in this figure, the experimental
data for all other $T$ also merged with this line. This scaling
behavior suggests that the Lorentzian distribution function for
log $t_{0}$ is intrinsic.\\
\indent Note that $F$(log $t_{0}$) follows not the Gaussian
distribution, but the Lorentzian distribution. For a statistically
independent random process, it is a basic statistical rule that
the resulting distribution should become Gaussian, regardless of
the process details \cite{Reif}. For example, impurities (or
crystal defects) inside a real crystal result in inhomogeneous
broadening of the light absorption line, which has a
Gaussian line shape.\\
\indent However, some studies have observed that magnetic
resonance line broadening of randomly distributed dipole
impurities follows the Lorentzian distribution \cite{Vleck}. The
first rigorous theoretical
 result for this problem is that of Anderson, who showed
that the distribution of any interaction field component in the
system of dilute aligned dipoles should be Lorentzian
\cite{Anderson, Klauder}. Polycrystalline FE films should contain
many dipole defects that will act as pinning sites for the domain
wall motion. To explain our observed Lorentzian distribution of
log $t_{0}$, we assume that a local field $\overline{E}$ exists at
the FE domain pinning sites and that it has a Lorentzian
distribution:
\begin{equation}
\textstyle F(\overline{E})=\frac{A}{\pi}
\left[\frac{\Delta}{\overline{E} ^{2}+ \Delta ^{2}}\right],
\end{equation}
where $\Delta$ is the half-width at half-maximum of the
$\overline{E}$ distribution function, related to the concentration
of
pinning sites.\\
\indent In the low $E_{ext}$ region, the domain wall motion should
be governed by thermal activation process at the pinning sites.
Without $\overline{E}$ effects, thermal activation results in a
domain wall speed in the form $v$ $\propto$ 1/$t_{0}$ $\propto$
$exp$[-($U$/$k_{B}T$)($E_{0}$/$E_{ext}$)], where $U$ is the energy
barrier and $E_{0}$ is the threshold electric field for pinned
domains \cite{Triscone}. Since $\overline{E}$ results in a change
in the effective electric field at pinning sites, the associated
$t_{0}$ can be expressed as
\begin{equation}
\textstyle t_{0} \sim
exp\left[(\frac{U}{k_{B}T})(\frac{E_{0}}{E_{ext}+\overline{E}})\right].
\end{equation}
Then, the distribution of $\overline{E}$ results in a distribution
in log $t_{0}$, using the relation
$F(log\,t_{0})=F(\overline{E})\cdot|d\overline{E}(log\,t_{0})/d(log\,t_{0})|$.
With
\begin{equation}
\textstyle log\,t_{1} \approx
\frac{UE_{0}}{k_{B}T}\cdot\left(\frac{1}{E_{ext}}\right)
\end{equation}
and \begin{equation} \textstyle w\,\approx \frac{U
E_{0}\Delta}{k_{B}T}\cdot\left(\frac{1}{E_{ext}^{2}}\right),
\end{equation}
we can obtain the desired Lorentzian distribution for $F$(log
$t_{0}$), i.e., Eq. (3), from Eqs. (4) and (5).

Our experimental values for log $t_{1}$ and $w$ agree with the
analytical forms. Figures 4(a) and (b) plot log $t_{1}$ $vs.$
1/$E_{ext}$ and $w$ $vs.$ 1/$E_{ext}^{2}$ at various $T$,
respectively. Both log $t_{1}$ and $w$ follow the expected
$E_{ext}$-dependence in the low $E_{ext}$ region. Note that Eq.
(6) is consistent with Merz's law \cite{Merz}, which states that
the current coming from FE polarization switching should have a
characteristic time of $exp$($\alpha$/$E_{ext}$), where $\alpha$
is the activation field. Using this empirical law, several studies
have measured $\alpha$ values. For example, So $et$ $al$. reported
$\alpha\approx$1700 kV/cm for 100-nm-thick epitaxial PZT films
\cite{YWSo}, and Scott $et$ $al$. reported $\alpha\approx$ 270
kV/cm for 350-nm-thick poly-PZT films \cite{Scott2}. These values
are consistent with our room temperature value of
$UE_{0}$/$k_{B}T$, i.e., 1400
kV/cm.\\
\begin{figure}[floatfix]
\includegraphics[width=2.2in]{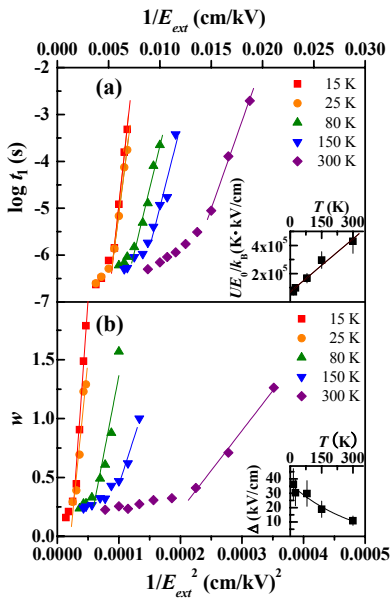}
\caption{(color online). $E_{ext}$-dependent (a) log $t_{1}$ and
(b) $w$ at various $T$. Note that log $t_{1}$ and $w$ are
proportional to 1/$E_{ext}$ and to 1/$E_{ext}^{2}$in the low
$E_{ext}$ region, respectively. The insets show $UE_{0}/k_{B}$ and
$\Delta$. The solid lines are guidelines for eyes.}
\end{figure}
\indent Our model viewed the FE domain switching kinetics as
domain wall motion driven by $E_{ext}$ with a random pinning
potential. In the low $E_{ext}$ region, thermal activation at the
pinning sites can be important, resulting in the so-called domain
wall creep motion. Applying atomic force microscopy, Tybell $et$
$al$. \cite{Triscone} and Paruch $et$ $al$. \cite{Paruch}
demonstrated that the domain-switching kinetics in epitaxial PZT
films are governed by the domain wall creep motion. Some
theoreticians studied the domain wall creep motion of an elastic
string in a random potential. They found a linear increase in $U$
with an increase in $T$ \cite{Kolton}. The insets in Fig. 4(a)
show that the value of $UE_{0}$/$k_{B}$ obtained from the linear
fits in Fig. 4(a) increase linearly with $T$, consistent with the
theoretical prediction for $U$ \cite{Kolton}. The inset in Fig.
4(b) shows $\Delta$ obtained from the fits to Fig. 4(b). Similar
exponential decay behavior was predicted in a magnetic
resonance study of randomly distributed dipoles \cite{Klein}.\\
\indent At this point, we wish to compare our model with the NLS
model. Although both models use Eq. (2), the origins and forms for
$F$(log $t_{0}$) are quite different. In the NLS model, the FE
film consists of numerous areas, each with its own and independent
$t_{0}$. Subsequently, it was suggested that the individually
switched regions correspond to single grains or clusters of grains
in which the grain boundaries act as frontiers limiting the
propagation of the switched region \cite{Stolichnov2}.
Consequently, the NLS model can be applied for polycrystalline
films only, and the form of $F$(log $t_{0}$) should depend on
their microstructure. Conversely, in our model, the interaction
between dipole defects inside the FE film induces a distribution
in the local field, which results in $F$(log $t_{0}$). Therefore,
both point defects and the grain boundaries could act as pinning
sites. Using the Lorentzian distribution for $F$(log $t_{0}$), our
model can be used for both epitaxial and polycrystalline FE films
\cite{YWSo}. Using Eqs. (2) and (3) with small $w$ values, we
could successfully explain the $\Delta P(t)$ for FE single
crystals or epitaxial thin films \cite{YWSo}. We also found that
our model can explain the $\Delta P(t)$ data for poly-PZT films
with Ti
concentrations of 0.48 and 0.65.\\
\indent Note that our model for thermally activated domain
switching kinetics can be viewed as the famous problem that treats
the propagation of elastic objects driven by an external force in
presence of a pinning potential \cite{Triscone,Paruch,Kolton}. It
can be applied to many FE films, since the domain wall motion with
a disordered pinning potential should be the dominant mechanism
for $\Delta P(t)$. Therefore, the $\Delta P(t)$ studies can be
used to investigate numerous intriguing issues concerning
nonlinear systems, such as creep motion, avalanche phenomenon,
pinning/depinning transition, and so on.\\
\indent In summary, we investigated the polarization switching
behaviors of (111)-oriented poly-PZT films and found that the
characteristic switching time obeyed the Lorentzian distribution.
We explained this intriguing phenomenon by introducing the local
electric field due to the defect dipole.\\
\indent The authors thank D. Kim for fruitful discussions. This
study was financially supported by Creative Research Initiatives
(Functionally Integrated Oxide Heterostructure) of MOST/KOSEF.


\begin{thebibliography}{30}
\bibitem{remark1} If all nuclei of opposite polarization arise
through whole process, $n$ could be larger than the actual
dimension.
\bibitem{remark2} Complications can occur due to charge
trapping or domain pinning, called the imprint effect. Refer to
Ref. \cite{Scott1}. To prevent the imprint effect, we applied a
pulse with the opposite polarity at the end of each pulse train
measurement (i.e., after pulse A2).
\bibitem{remark3} A double exponential function
$exp$[-\{10$^{log\,t}$/10$^{log\,t_{0}}$\}$^{n}$] with $n$ $>$1
can be approximated as a step function centered at
log$\;t_{0}$=log$\;t$. As a result, Eq. (2) can be approximated as
$2P_{s}A/\pi\cdot [arctan\{(log\,t-log\,t_{1})/w\}+\pi/2]$.
\end{thebibliography}
\end{document}